\documentstyle[prl,aps,epsf]{revtex}\def\narrowtext{}\tighten\twocolumn
\input epsf.sty
\begin{document}
\draft

\title{
Slow transverse relaxation
of the Zn-neighbor Cu nuclear spins
in YBa$_2$(Cu$_{1-x}$Zn$_x$)$_4$O$_{8}$
}
\author{
Y. Itoh 
}

\address{
Department of Chemistry, Graduate School of Science,\\
Kyoto University, Kyoto 606-8502, Japan \\ 
}

\date{\today}%
\maketitle %

\begin{abstract} 
An explanation is given for the recent observation 
in the plane-site Cu nuclear quadrupole resonance (NQR) study of 
a heavily Zn-substituted YBa$_2$(Cu$_{1-x}$Zn$_x$)$_4$O$_{8}$
that the Cu nuclear spin-spin relaxation time $T_{\mathrm 2G}$ near the impurity Zn 
is longer than that away from Zn.
The long $T_{\mathrm 2G}$ of the Zn-neighbor Cu nuclear spins results from 
a distant coupling between only a few nuclear spins near Zn,
which does not indicate the absence of local enhancement of the staggered spin susceptibility
around Zn.  
\end{abstract}

\narrowtext 

The Zn-induced local magnetic correlation in impurity Zn-substituted high-$T_{\rm c}$ superconductors
has attracted great interests and brought about a lot of debates.
In Refs.[1,2], it was debated whether the local magnetic correlation near Zn 
is enhanced or suppressed. 
Recently, in Refs.[3,4], it was debated whether the local staggered spin susceptibility 
is enhanced near Zn or literal local moments are induced. 

In Ref.[4], Williams {\it et al}. criticized our Cu NQR study~\cite{ItohZn} 
for lightly Zn-substituted YBa$_2$(Cu$_{1-x}$Zn$_x$)$_4$O$_{8}$ (Y1248),
using their NQR results for heavily Zn-substituted samples. 
Williams {\it et al}. criticized the site assignment of the satellite Cu NQR signals 
to the fourth nearest neighbor Cu sites about Zn. 
Actually, we assign the satellite signals 
to the {\it fraction} of the fourth nearest neighbor Cu sites in Y1248~\cite{ItohZn}. 
They assigned the satellite NQR signals to the first nearest neighbor Cu sites
and claimed that the local magnetic correlation near Zn is not enhanced,
because they observed a weak frequency dependence of Cu nuclear spin-lattice relaxation time $T_1$
and a longer Cu nuclear spin-spin relaxation time $T_{\mathrm 2G}$ at the satellite than that at the main NQR signals.
 
We have already reported a nonmonotonic change in the spin correlation 
from lightly to heavily Zn-substitution regimes~\cite{ItohMIT}
and the strong frequency dependence of $T_1$ for the lightly Zn-substituted Y1248~\cite{ItohZn}.
Already-existing reports indicate that the absence of rapid enhancement of transverse relaxation rates
of the observable nuclear spins 
is the characteristic of the wipeout effect on Zn-substituted high-$T_{\rm c}$ superconductors
~\cite{ImaiZn,Kobayashi,Yamagata1,Yamagata2}.
The wipeout effect results from primarily 
a low frequency anomaly of the dynamical spin susceptibility related to the $T_1$ process
but not necessarily a high frequency anomaly related to the static staggered spin susceptibility and $T_{\mathrm 2G}$.
For the heavily Zn-substituted Y1248, the antiferromagnetic network
on the CuO$_2$ plane is partly disconnected by the nonmagnetic impurities, 
so that the local magnetic correlation is also suppressed. 
It is likely that in the heavily Zn-substituted Y1248, a part of Zn impurities is
substituted for the chain Cu site, 
because we have already observed the partial substitution effect
by the chain-site Cu NQR experiments (not shown here). 
Technically, the satellite and the main Cu NQR spectra may be so broad
and indistinguishable in the frequency dependence of $T_1$ for the heavily Zn-substituted Y1248~\cite{WK}. 

Although Williams {\it et al.}~\cite{Williams} criticize our model using binomial distribution functions~\cite{ItohZn} 
for being "a complicated statistical model"
and     
the impurity-induced NMR relaxation theory for being "a nonunique model,"
these criticisms are irrelevant, because
the model and the theory have given satisfactory results 
for dilute alloys and dilute Heisenberg systems so far.  
Shortcomings of these analyses for the high-$T_{\rm c}$ superconductors as well as the quantum spin systems 
have already been recognized 
and totally discussed before.  

In this paper, we focus on the unsolved problem in~\cite{WK}; 
the reason why the slow nuclear spin-spin relaxation 
at the satellite Cu NQR lines was observed in the Zn-substituted Y1248
and how it is compatible with the local enhancement of staggered spin susceptibility around Zn.
One should note that the literal local moments induced at the first nearest neighbor  Cu site about Zn~\cite{Williams}
is not compatible with 
the slow nuclear spin-echo decay~\cite{WK}  
even if the absence of the local enhancement of the staggered spin susceptibility.
Because the $I_z$-fluctuation due to a short $T_1$ and the Curie magnetism 
must enhance the nuclear spin-echo decay rate.

First, let us review how the Cu nuclear spin-spin relation rate 1/$T_{\mathrm 2G}$
is expressed by the staggered spin susceptibility $\chi({\bf Q})$ (${\bf Q}$=[$\pi$, $\pi$])
in the high-$T_{\rm c}$ superconductors.
The Cu nuclear spin-spin Hamiltonian~\cite{PS1,PS2} is expressed by 
\begin{equation}
H_{II}=\sum_{i\neq j}^{} a_{ij}I_{iz}I_{jz}.
\label{e.NSS}
\end{equation}
Here, we neglect the effect of $I_z$-fluctuation by a short $T_1$~\cite{Curro}.  
A range function $a_{ij}$ between the nuclear spins $I_{i}$ and $I_{j}$~\cite{PS2,Itoh1} is given by  
\begin{equation}
a_{ij}=\int d\mathrm{\bf q}{A}^{2}(\bf q)\mathrm{Re}\chi({\bf q})\mathrm{exp}(i{\bf q}\cdot {\bf r}_{ij}),
\label{e.Range1}
\end{equation}
where $A({\bf q})$ is the Fourier transform of the hyperfine coupling constants
between a nuclear and an electron spins. 
The high-$T_{\rm c}$ superconductors are considered to be
quasi-two dimensional, nearly antiferromagnetic electron systems.
The static spin susceptibility $\chi({\bf q})$
is enhanced at or around the antiferromagnetic wave vector ${\bf q}$=${\bf Q}$. 
Using the dynamical spin susceptibility $\chi({\bf q}, \omega)$ in~\cite{MMP,MTU},
we obtain the range function~\cite{Itoh1} 

\begin{equation}
a_{ij}\propto A^{2}({\bf Q})\chi_0({\bf Q})\sqrt\frac{\xi}{{\it r}_{ij}}\mathrm{exp}(-\frac{{\it r}_{ij}}{\xi}),
\label{e.Range2}
\end{equation} 
where $\xi$ is an antiferromagnetic correlation length, 
$\chi_{0}({\bf Q})$ is the staggered spin susceptibility of bare electron gas, 
and $\chi({\bf Q})$=$\chi_{0}({\bf Q})(\xi/q_{\mathrm B})^2$/$\alpha_sA$ 
($q_{\mathrm B}$ is the cut off wave vector, 
$\alpha_s$ is the exchange enhancement factor,
and $A$ is $|\partial^2 \chi_{0}({\bf q})/\partial {\bf q}^2|_{\bf q=Q}/\chi_{0}({\bf Q})$) 
is given in~\cite{MTU}.
For Y1248, we estimated $\xi\sim$3.4$a$ ($a$ is an in-plane lattice constant) at 100 K~\cite{ItohSCR,ItohAFL}.  

For simplicity, we assume a nuclear spin $I$=1/2 system. 
For the nuclear spins away from the impurities, 
the nuclear spin-spin relaxation curve
$E(2\tau)$  as a function of time $\tau$ in a sequence of $\pi$/2-$\tau$-$\pi$ pulses
is given by
\begin{eqnarray}
E(2\tau)&=&E_{0}\prod_{j=1}^{} \mathrm{cos}(2{\it a}_{ij}\tau)\\
&\approx&E_{0}\mathrm{exp}[-\frac{(2\tau)^2}{2}\sum_{j=1}^{}{\it a}_{ij}^2], 
\label{e.Decay}
\end{eqnarray}
~\cite{PS1,Slichter1,Slichter,Abragam}.  
Using eq.(2) or (3), we obtain the square of the Gaussian relaxation rate of nuclear spin-echo decay
\begin{equation}
\Bigl(\frac{1}{T_{\mathrm{2G}}}\Bigr)^2=\sum_{j=1}^{\infty}a_{0j}^2\propto{A}^{4}(\bf Q)\mathrm{Re}\chi({\bf Q})^2, 
\label{e.Gaussian}
\end{equation}
~\cite{Itoh1,Itoh2,Thelen,Takigawa}.  
One should note that eq. (6) is not applicable to a finite spin system,
in which a few nuclear spins are distantly coupled with each other.
e.g. the first and the fourth nearest neighbor Cu spins around Zn.
The satellite Cu NQR corresponds to such a system.  
Unfortunately, Williams {\it et al}. applied eq. (6) to the satellite Cu nuclear spin-echo decay~\cite{Williams,WK}.
 
For the first nearest neighbor Cu nuclei around Zn, we obtain

\begin{equation}
\Bigl(\frac{1}{^{\mathrm{1st}}T_{\mathrm{2G}}}\Bigr)^2\approx\sum_{k=1}^{2}a_{1k}^2,
\label{e.T2G1stnn}
\end{equation}
where $k$ belongs to the 1st nearest neighbor Cu sites around Zn.
Figure 1 illustrates the CuO$_2$ plane with an impurity Zn.
For the fourth nearest neighbor Cu nuclei around Zn, we obtain

\begin{equation}
\Bigl(\frac{1}{^{\mathrm{4th}}T_{\mathrm{2G}}}\Bigr)^2\approx\sum_{l=1}^{6}a_{4l}^2,
\label{e.T2G4thnn}
\end{equation}
where $l$ belongs to the 4th nearest neighbor Cu sites around Zn.  

Not $\chi({\bf Q})$ in eq. (6) but eq. (3) should be taken into consideration
for the first and the fourth nearest neighbor Cu nuclear spins. 
Equations (7) and (8) are the finite lattice sums
of the square of the range function of eq. (3). 
In rough estimation of eqs. (7) and (8) from eq. (3),
since $\xi\sim$3.4$a$ and assuming $a_{0j}\approx a_{1j}\approx a_{4j}$, 
then about 36 nuclear spins contribute eq. (6), 
whereas only 2 spins contribute eq. (7) and 
about 6 or less spins contribute eq. (8). 
Thus, we obatin 
\begin{equation} 
\Bigl(\frac{1}{T_{\mathrm{2G}}}\Bigr)^2>>\Bigl(\frac{1}{^{\mathrm{1st}}T_{\mathrm{2G}}}\Bigr)^2,
\Bigl(\frac{1}{^{\mathrm{4th}}T_{\mathrm{2G}}}\Bigr)^2.
\label{e.FiniteT2G}
\end{equation} 

If the satellite Cu NQR signals are assigned to the first nearest neighbor Cu sites
as in~\cite{Williams},
the spin-echo modulation of cos$^2$(2$a_{k1}\tau$) must be observed
but has never been reported.
Thus, the slow nuclear spin-spin relaxation does not indicate
the absence of the enhancement of the local staggered spin susceptibility around Zn
but rather supports
our site assignment of the satellite Cu NQR signals 
to a part of the fourth nearest neighbor Cu sites~\cite{ItohZn}
even for the heavily Zn-substituted Y1248. 

Eventually, the problem is how much the local staggered spin susceptibility
in $a_{4l}$ and $a_{1k}$
is enhanced around Zn. 
The detailed measurements of $T_{\mathrm 2G}$ 
of the satellite Cu NQR signals for the lightly Zn-substituted Y1248
might be a theoretical constraint on modeling the staggered susceptibility about Zn.   
 
\acknowledgments 
I would like to thank T. Machi and K. Yoshimura 
for their stimulating discussions.
This study was supported by a Grant-in-Aid for Science Research on 
Priority Area,
'Invention of anomalous quantum materials' from the Ministry of 
Education, Science, Sports and Culture of Japan (Grant No. 16076210).

\begin{figure}
\epsfxsize=2.4in
\epsfbox{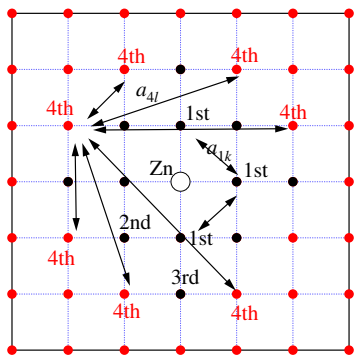}
\vspace{0.0cm}
\caption{
Top view of a CuO$_2$ plane with a Zn impurity (an open circle).
Closed circles are the Cu ions. Oxygen ions are omitted. 
The arrows indicate the pairs of nuclear spin-spin coupling.
$a_{1k}$ ($k$=1, 2) is the coupling constants between the first nearest neighbor Cu nuclear spins about Zn
and $a_{4l}$ ($l$=1, 2, $\cdots$, 6) is those between the fourth nearest neighbor Cu nuclear spins about Zn.
}
\label{CuNMR}
\end{figure}

\end{document}